# Performance Improvement of LTS Undulators for Synchrotron Light Sources

Emanuela Barzi, *Senior Member, IEEE*, Masaki Takeuchi, Daniele Turrioni, and Akihiro Kikuchi

*Abstract*— The joint expertise of ANL and FNAL has led to the production of $Nb_3Sn$ undulator magnets in operation in the ANL Advanced Photon Source (APS). These magnets showed performance reproducibility close to the short sample limit, and a design field increase of 20% at 820A. However, the long training did not allow obtaining the expected 50% increase of the on-axis magnetic field with respect to the ~1 T produced at 450 A current in the ANL NbTi undulator. To address this, 10-pole long undulator prototypes were fabricated, and CTD-101K® was replaced as impregnation material with TELENE®, an organic olefin-based thermosetting dicyclopentadiene resin produced by RIMTEC Corporation, Japan. Training and magnet retraining after a thermal cycle were nearly eliminated, with only a couple of quenches needed before reaching short sample limit at over 1,100 A. TELENE will enable operation of $Nb_3Sn$ undulators much closer to their short sample limit, expanding the energy range and brightness intensity of light sources. TELENE is Co-60 gamma radiation resistant up to 7-8 MGy, and therefore already applicable to impregnate planar, helical and universal devices operating in lower radiation environments than high energy colliders.

## I. INTRODUCTION

IN the past four decades, third generation light sources have operated and provided low emittance beams thanks to insertion devices, i.e. undulators, that produced high brightness beamlines [1]. There are now dozens of light sources in the world. Undulators deliver synchrotron radiation in narrowband beams, and their peak magnetic field is tuneable by changing the gap between the top and bottom arrays. Until recently, permanent magnets were used to supply fundamental and multiple odd harmonics with short periods and sufficient on-axis magnetic field strength. However, even in-vacuum cryogenically cooled undulators made with permanent magnet technology have reached their limit in the achievable on-axis peak magnetic field for a given period length and magnetic gap [2].

Superconducting undulators (SCU) are therefore needed to keep improving undulator performance, i.e. shorter periods and higher on-axis peak fields. Associated challenges include cryogenic and mechanical design that allows operation with cryocoolers, and long-term performance in a radiation environment [1]. To beat these challenges, Argonne National Laboratory (ANL) established at the Advanced Photon Source (APS) a dedicated facility, which included a magnet winding machine, an epoxy impregnation setup, liquid helium testing cryostats, and special cryostats operating with cryocoolers [2]. The ANL SCU undulators magnets currently operating at the APS are indirectly cooled with liquid helium penetrating through channels in the magnet cores. The liquid helium is stored in a tank cooled by cryocoolers.

TABLE I
SCU PLANAR UNDULATOR SPECIFICATIONS [3, 14]

| Undulator Specifications | $Nb_3Sn$ | NbTi |
|---|---|---|
| On axis design field, T | 1.17 | 0.97 |
| ~K value | 2 | 1.6 |
| Design current, A | 820 | 450 |
| Period length, mm | 18 | 18 |
| Magnetic gap, mm | 9.5 | 9.5 |
| Magnetic length, m | 1.1 | 1.1 |

The first SCU planar undulator designed and fabricated at the APS was made of NbTi [2]. Its specifications are shown in Table I. This magnet was impregnated with CTD-101K. Its typical performance and training behavior is in Fig. 1 [3].

The natural next step for the APS was to develop a $Nb_3Sn$ SCU. This is because the NbTi undulator magnet technology is limited in smaller period values. With $Nb_3Sn$, a smaller period can be produced for the same magnetic field and magnetic gap, than with NbTi. Within a SLAC National Accelerator Laboratory, ANL and Lawrence Berkeley National Laboratory (LBNL) collaboration from 2014 to 2016 [4], LBNL produced a 1.5 m long $Nb_3Sn$ undulator [5] of 19 mm period, 8 mm magnetic gap and K value of 3.2. Its performance and training behavior is in Fig. 1.

Then in 2018, ANL started a new collaboration with Fermi National Accelerator Laboratory (FNAL) and LBNL, funded by the DOE BES accelerator R&D program, with the goal of producing a $Nb_3Sn$ undulator to be installed in the APS storage ring for user operations [6]. FNAL High Field Magnet Program (now part of the US Magnet Development Program) has been

Paper submitted on xy date. Work supported by Fermi Research Alliance, LLC, under contract No. DE-AC02-07CH11359 with the U.S. DOE. Corresponding author: Emanuela Barzi.

Emanuela Barzi and Daniele Turrioni are with Fermi National Accelerator Laboratory, Batavia, IL 60510, USA (e-mails: barzi@fnal.gov, turrioni@fnal.gov).

Masaki Takeuchi is with RIMTEC Corporation, Kurashiki-shi, Okayama 711-0934, Japan (e-mail: m5.takeuchi@rimtec.co.jp).

Akihiro Kikuchi is with the National Institute for Materials Science, Tsukuba, Ibaraki 305-0047, Japan (e-mail: kikuchi.akihiro@nims.go.jp).

developing Nb$_3$Sn superconducting magnets, materials and technologies for present and future particle accelerators since the late 1990s, culminating with its 2020 world record field of 14.6 T for accelerator dipoles [7]. One necessary key factor that was learned at FNAL to insure Nb$_3$Sn magnet performance is that of preventing magnetic or flux jump instability of the superconductor being used in the coils. Magnetic instability at low magnetic fields can prevent a superconducting magnet from reaching its design current level. From [8] it was determined that the sub-element size of a Nb$_3$Sn wire with J$_c$ (4.2K, 12 T) of 2800 A/mm$^2$ had to be less than 36 μm for stable operation. Therefore, a Restacked Rod Processed (RRP) wire of 0.6 mm in diameter, and with 144 superconducting subelements over 169 total subelements was used. Its equivalent subelement diameter was ~35 μm, and the critical current density J$_c$ (4.2K, 12 T) was ~2500 A/mm$^2$. The first phase of the project focused on small undulator models of just nine racetrack coils wound in a groove between ten poles [9, 10]. There were 46 turns in each groove, and each period included two grooves and two poles. The S2-glass braided Nb$_3$Sn wire was continuously wound turn-by-turn between the poles. The magnets were designed and fabricated at ANL with FNAL input, heat treated in argon atmosphere at FNAL, and impregnated with epoxy and tested at ANL. This first phase insured establishing a sound magnet design and optimizing the high temperature heat treatment process to increase the stability margin with respect to the operation current [11]. The scale-up of the magnets to 0.5 m occurred in the second phase [12, 13], which culminated in the fabrication and testing of a 1.1 m long Nb$_3$Sn undulator magnet [14], which is presently in operation at the APS. Its performance and training behavior is in Fig. 1.

In this paper we will show how the use of TELENE as impregnation material allows operation currents much closer to a magnet's short sample limit, thanks to training elimination. This alone makes superconducting magnets much more cost-effective.

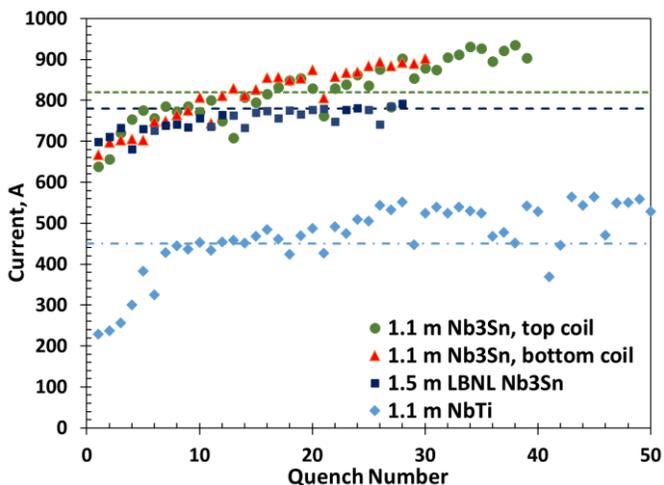

**Fig. 1.** Performance and training behavior of scaled-up superconducting undulators [3, 5, 14] developed for ANL APS, compared with the respective operation currents, shown as dashed lines.

## II. RESULTS AND METHODS FOR PURE TELENE

Long training has been a feature of NbTi and Nb$_3$Sn impregnated magnets for decades. Energy deposition that initiates quenches can emanate from a variety of sources (magnetic flux jumps, conductor motion, epoxy cracking, materials' interfaces, etc.). All these sources contribute to a resulting "disturbance spectrum". Any attempt made so far to reduce magnet training with materials and methods applicable to accelerator magnets failed. We present here results in improving the training behavior in Nb$_3$Sn undulator magnets using as coil impregnation material C$_{10}$H$_{12}$, an organic olefin-based thermosetting dicyclopentadiene (DCP) resin, in replacement of the CTD-101K® epoxy currently used for this purpose. This resin is commercially available as TELENE® by RIMTEC Corporation, Japan. TELENE was chosen for these studies because of the following main reasons: 1. Its ductility, i.e. the ability to accept large strains; 2. Its toughness, i.e. the amount of energy per unit volume that the material can absorb before rupturing, or the area underneath the stress vs. strain curve; 3. Its potential for radiation resistance. Also, TELENE's pot life of up to 3.5 hours at 5°C ensures scalability to impregnate larger coil volumes [15].

TELENE was used to impregnate ANL Nb$_3$Sn short undulator models, which were fabricated at ANL with the same S2-glass braided Nb$_3$Sn wire of 0.6 mm as used in [6]. After winding, the magnets were heat treated at FNAL in argon atmosphere using well-established treatment cycles. Then they were placed in a leak-tight impregnation mold for TELENE impregnation either at ANL or at FNAL.

TELENE impregnated magnets were tested at FNAL at 4.2 K in liquid helium, in a cryostat of the Superconducting R&D lab, by using an insert equipped with 2000 A DC leads. Two pairs of voltage taps, each covering half of the magnet, were used. The voltage tap wires were connected to an NI-9239 card of a compact RIO DAQ system. The NI card had 4 channels with an acquisition frequency of 50 kHz and 24 bits per channel. The threshold for the quench protection system was 100 mV for the differential voltage. To diagnose the origins of the few quenches observed in the second and third run after achieving short sample limit with the first Nb$_3$Sn tested undulator (Figs. 2 and 5), advanced instrumentation was added to the coils, including quench antennae (QA) and acoustic sensors on each end of the coil. They were read with an 8-channel 20 GHz oscilloscope. A new QA system, specific to the Nb$_3$Sn undulator magnet model, will be designed and fabricated, and analysis of QA data will be performed by FNAL experts.

The first TELENE impregnated undulator magnet achieved short sample limit (SSL) at every training cycle after only two quenches, compared with ~100 when CTD-101K was used on a number of identical undulator coils [15]. The quench data from [15] for the first TELENE impregnated short undulator model as compared with identical models impregnated with CTD-101K from [6] are shown in Fig. 2 for the convenience of the reader. As can be seen, the Nb$_3$Sn undulator magnet models impregnated with CTD-101K performed reproducibly starting with model SMM3. However, because of long training, the expected short sample limit was never truly attained, and the operation current was set at 820 A, as also shown in Fig. 1 for the full-scale magnet. As can be seen from Fig. 3, this current produces only a 20% increase of the on-axis magnetic field with respect to the ~1 T obtained at 450 A in the ANL NbTi undulator version. When replacing CTD-101K with TELENE as impregnation material, training and magnet retraining were nearly eliminated before reaching short sample limit at over 1,100 A. Training elimination allows reducing the operation margin of these magnets. In this specific case, the operation current can be increased to at least 1000 A, which corresponds to an on-axis maximum magnetic field of ~1.4 T, i.e. a 40% increase with respect to the NbTi undulator magnets (see Fig. 3).

TELENE can therefore enable operation of $Nb_3Sn$ undulators much closer to their short sample limit, expanding the energy range and brightness intensity of light sources. Pure TELENE is Co-60 gamma radiation resistant up to 7-8 MGy, and therefore already applicable for impregnation of insertion devices for synchrotron light sources, operating in lower radiation environments than high energy colliders.

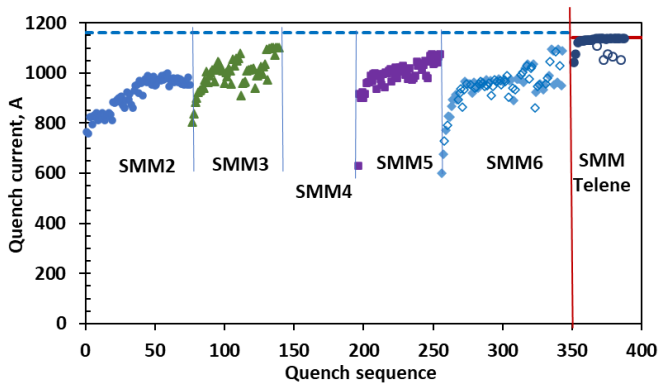

**Fig. 2.** Quench history of TELENE impregnated short undulator model as compared with that of nearly identical undulator short models impregnated with CTD-101K [10, 15]. The two dashed lines represent the short sample limits for the epoxy (1160 A) and TELENE (1143 A) impregnated ones respectively.

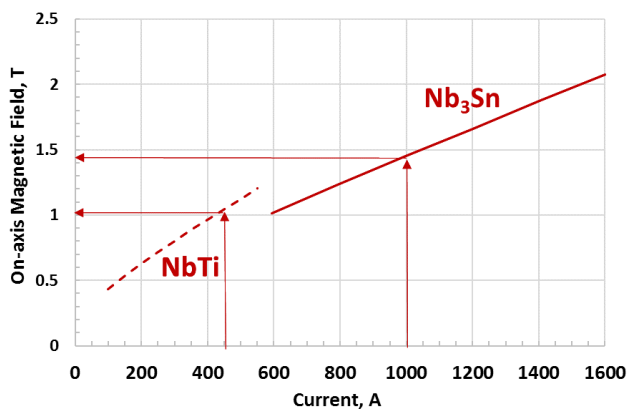

**Fig. 3.** On-axis peak magnetic field as function of operational current for the SCU undulators [3, 14] in Table I.

TELENE was successful also in preventing quenching up to at least 95% of its short sample limit in a NbTi dipole made of 6/1 round cable. This was done by adapting an original superconducting transformer designed for Rutherford cable tests up to 25 kA and 15 T of external field to testing small dipole coils of 0.35 m of maximum length. The NbTi dipole was made of a round cable with 6 NbTi strands of 0.8 mm wrapped around a Cu wire also of 0.8 mm. It was impregnated with pure TELENE and reached ~6000 A, i.e. 95% of its short sample limit without quenching. The test system was limited at 6000 A due to the larger inductance of the tested coil as opposed to the original bifilar configuration used when testing Rutherford cable. This experimental setup will be upgraded in the close future to achieve at least 10,000 A of current in actual coils. This additional successful result makes TELENE applicable to reducing or eliminating training also in NbTi undulators.

In addition to undulator magnets for storage rings, another motivating application for TELENE is for X-ray Free Electron Lasers (FELs), since superconducting undulators are ideal for helical magnetic field configurations with smaller period. "Superconducting undulators can extend the spectral range of existing FEL sources and can fit in shorter tunnels without degrading FEL performance." [2] Within the initiative described in [4], as a FEL superconducting magnet prototype, ANL designed and produced a 1.5 m NbTi planar undulator which met all technical specifications [16]. Implementing TELENE to this type of magnets will further improve their performance and training behavior.

In short, TELENE will allow exploiting any superconducting undulator magnet for light sources to the fullest of the conductor capabilities and reduce operation margins.

### III. RESULTS AND METHODS FOR MIXED TELENE

The potential of improving TELENE's thermal properties by mixing it with high-specific heat ($C_p$) ceramic powders was another component of this research. Several mixed resins were produced in [15]. TELENE was mixed with 45wt%, 61wt%, and 82wt% of $Gd_2O_3$ powder (0.7 to 1.2 μm in size); with 21.5wt% to 87wt% of $Gd_2O_2S$ powder; and with 83wt% of $HoCu_2$ powder. The viscosity, $C_p$, thermal conductivity (k), and other physical properties of these mixed resins were measured as function of temperature and magnetic field. The TELENE-87wt%$Gd_2O_2S$ had a peak in $C_p$ between 4.3K and 5.3K at fields between 0 and 8 T and the largest thermal conductivity over the whole temperature range [15].

We have investigated radiation resistance of pure and mixed TELENE resins after irradiating samples up to 10 MGy+ using a Co-60 gamma-ray irradiation facility at the Takasaki Advanced Radiation Research Institute in Japan. TELENE-87wt%$Gd_2O_2S$ also had the best radiation resistance. However, it does not feature the same desirable mechanical properties of pure TELENE [15]. In Fig. 4, we show the flexural stress as function of strain for TELENE resins in which the wt% of the $Gd_2O_2S$ was reduced to 43% and to 21.5% to preserve their ductility. These new data are compared with those from [15] in the plot of Fig. 4.

Based on these results, TELENE-43wt%$Gd_2O_2S$ was selected to impregnate a second $Nb_3Sn$ undulator short model. The quench history results of this second undulator magnet are shown in Fig. 5. As a rule of thumb, with pure TELENE, training is achieved with a couple of quenches; with TELENE-43wt%$Gd_2O_2S$ with 20 quenches; and with CTD-101K with 100 quenches. A reason for the TELENE-43wt%$Gd_2O_2S$ to be less effective than pure TELENE in eliminating training is its much lower thermal diffusivity $D = k/(\rho\ C_p)$ than for pure TELENE, due to its larger $C_p$ ($\rho$ is the material's density.) To make use of the radiation resistant properties of TELENE mixed with ceramic powders, thermal conductivity of these resins needs to be increased through materials engineering, i.e. by adding high-thermal conductivity components in their composition.

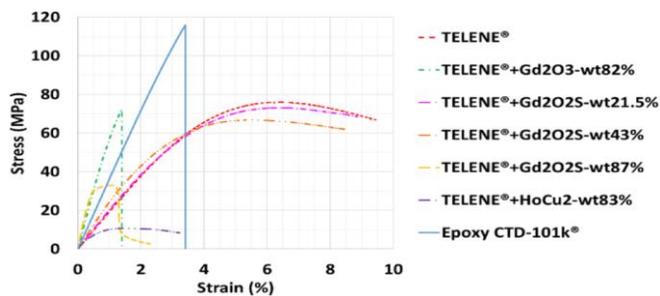

**Fig. 4.** Flexural stress vs. strain curves for TELENE-21.5wt%$Gd_2O_2S$ and TELENE-43wt%$Gd_2O_2S$, as compared with pure and mixed TELENE resins from [15].

To increase thermal diffusivity, NIMS and RIMTEC have produced new TELENE resins with higher thermal conductivity by mixing them with Aluminum Nitride (AlN), which is an excellent material to use if high thermal conductivity and electrical insulation properties are both desired. New resins were fabricated by combining the powder fillers and the TELENE using a planetary mixer. The viscosity of the resins is controlled by the volume fraction and average size of the powder filler, and is measured with a type- B viscometer. To produce the new resins, TELENE was mixed with 43wt%, 53wt%, 63wt%, 79wt%, and 84wt% of AlN powder (1 to 20 μm in size); and with 31wt% of AlN and 43wt% of $Gd_2O_2S$ powder. The chemical composition, average powder size and volume fraction are still being optimized.

The minimum quench energy (MQE) of impregnated NbTi wires is measured on ITER-type barrels. Two strain gauges are used as heaters for each sample, and glued to it using STYCAST 2850FT. The instrumentation wires are soldered before sample and strain gauges get brushed with a mixed TELENE resin. A 200 W power supply provides the excitation voltage to the strain gauges. Using a LabView DAQ program, a pulse output is generated from the power supply and the voltage across the strain gauge is measured. With the $I_c$ of the sample first measured, a constant bias current below $I_c$ is applied to the sample while heat pulses are fired using the strain gauge. A separate quench protection system monitors the voltage across the sample and shuts down the power supply if the quench threshold is reached. By gradually increasing the pulse energy, the minimum energy that induces a quench is defined as the MQE of the sample [15].

The MQE of the impregnated 0.8 mm NbTi wire samples was measured for heater pulse durations from 200 ms to 1.5 s, with $I_c$% of up to 90% and magnetic fields between 6 and 9 T. At 9 T, the $I_c$(4.2 K) was 140 A. Samples of TELENE-31wt%AlN-43wt%$Gd_2O_2S$, TELENE-43wt%AlN (1 μm powder) and TELENE-84wt%AlN (20 μm powder) were tested. In Fig. 6, we show the MQE, obtained at 9 T and at 80% of $I_c$ as function of pulse duration, for these resins. These new data are compared with those from [15] in the plot of Fig. 6. The high thermal conductivity resins show larger increases in MQE as a function of pulse time than the high specific heat resins.

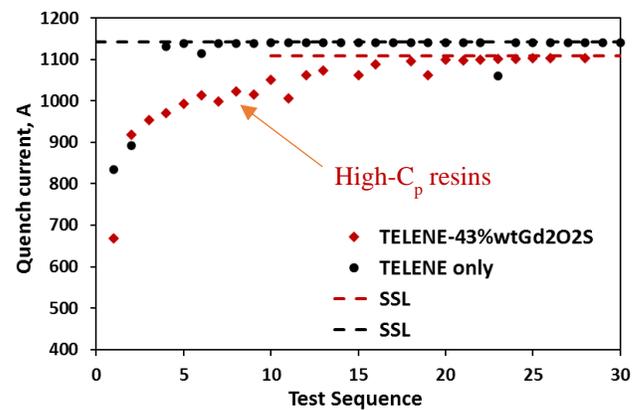

**Fig. 5.** Quench history comparison of first $Nb_3Sn$ undulator model impregnated with pure TELENE and with TELENE-43wt%$Gd_2O_2S$.

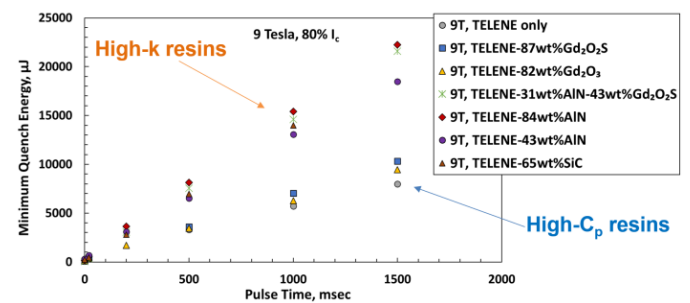

**Fig. 6.** Minimum Quench Energy vs. heater pulse duration at 80% of the critical current $I_c$ at 9 T for NbTi wire samples impregnated with TELENE-31wt%AlN-43wt%$Gd_2O_2S$, TELENE-43wt%AlN (1 μm powder) and TELENE-84wt%AlN (20 μm powder), as compared with pure and mixed TELENE resins from [15].

## V. CONCLUSION

The $Nb_3Sn$ undulator magnet models originally developed by ANL and FNAL, when impregnated with CTD-101K performed reproducibly close to the short sample limit and with an on-axis magnetic field increase of 20% at 820 A. This is because the long training did not allow obtaining the expected 50% increase with respect to the ~1 T produced at 450 A current in the ANL NbTi undulator version. When replacing CTD-101K with TELENE as impregnation material, training and magnet retraining were nearly eliminated before reaching short sample limit at over 1,100 A. By operating at 1000 A, an on-axis maximum magnetic field of ~1.4 T is produced, which is a 40% increase with respect to the NbTi undulator magnets. TELENE will therefore enable operation of $Nb_3Sn$ undulators much closer to their short sample limit, expanding the energy range and brightness intensity of light sources.

TELENE was successful in preventing quenching up to 95% of its short sample limit (because of the test system current limits) in a NbTi accelerator dipole. This makes TELENE applicable to reduce or eliminate training also in NbTi undulators.

In addition to undulator magnets for storage rings, another inspiring application for TELENE is for X-ray Free Electron Lasers (FELs). Implementing TELENE to this type of magnets will further improve their performance and training behavior. In short, TELENE will allow exploiting any superconducting undulator magnet for light sources to the fullest of the conductor capabilities and reduce operation margins.

High-$C_p$ ceramic powders mixed in TELENE have proven to be exceptionally radiation resistant to Co-60 gamma irradiation. When combined with the ductility and toughness properties of TELENE, these resins have already shown superior training performance with respect to CTD-101K. To fully exploit their characteristics, their thermal diffusivity D was increased by mixing also high-thermal conductivity components in these resins.

TELENE was successful in preventing training in the $Nb_3Sn$ ANL undulator, which produces a maximum magnetic field of about 5 T and maximum equivalent stress on the conductor of less than 100 MPa. Our next step is to check whether the developed resins can lead also to a reduction in training in stress managed accelerator magnets, which is the current core design in the US Magnet Development Program (US-MDP).

ACKNOWLEDGMENT

E. Barzi thanks Alexander Zlobin, from FNAL, who gave her the idea for this paper. The authors also thank Ibrahim Kesgin, from Argonne National Laboratory, for winding the small undulator models used in this research, as well as for designing and providing the epoxy impregnation mold. The authors are looking forward to working with the rest of our US and Japanese collaboration members Joe Di Marco (FNAL), Ibrahim Kesgin (ANL PI), Diego Arbelaez (LBNL PI), Mark Palmer (BNL PI), Tatsushi Nakamoto (KEK PI), Xudong Wang (KEK), Yasuo Iijima (NIMS), and Kazuto Hirata (NIMS), for the 2-year renewal of the LAB 24-3200 U.S.-Japan Science and Technology Cooperation Program in HEP SCIENCE AND TECHNOLOGY COOPERATION PROGRAM IN HIGH ENERGY titled "High heat capacity and radiation-resistant thermally conducting organic resins for impregnation of high field superconducting magnets."

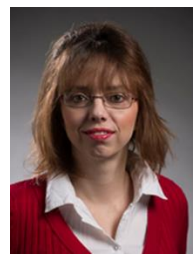

**Emanuela Barzi** (Senior Member, IEEE) is a senior scientist at Fermi National Accelerator Laboratory and an adjunct professor at Ohio State University. A 2012 Fellow of the American Physical Society (APS), a 2021 Fellow of the International Association of Advanced Materials, and a senior member of the IEEE, Barzi has been an active member of the high-energy accelerator and physics communities for nearly 30 years. The Superconducting R&D lab that she founded at FNAL is a world leading research center in low- and high-temperature superconductor technologies for the next generation of particle accelerators. Barzi is a member of the FNAL team that in 2020 produced a world-record field of 14.6 Tesla for a $Nb_3Sn$ accelerator dipole magnet, is co-leading the multi-lab effort on $Nb_3Sn$ undulators for ANL Advanced Photon Source, is FNAL coordinator of five trilateral EU-US-Japan collaborations, and is a member of the Muon g-2 Collaboration. She has co-authored more than 260 peer-reviewed papers and book chapters with close to 9000 citations. In 2010 she was awarded the Japanese "Superconductor Science and Technology Prize." Barzi also established extensive educational programs at FNAL for graduate students in Physics and Engineering, including the Italian Graduate Students Program at FNAL, that have benefited hundreds of young professionals, and has mentored 40+ students in her lab for internships, Masters and PhDs. She is a former councilor of the APS Forum on International Physics, and a former member of the APS Council Steering Committee and of the APS Ethics Committee, among others.

**Masaki Takeuchi**, photograph and biography not available at the time of publication.

**Daniele Turrioni**, photograph and biography not available at the time of publication.

**Akihiro Kikuchi**, photograph and biography not available at the time of publication.